\documentclass[%
 reprint,
 superscriptaddress,
 amsmath,amssymb,
 aps,
 pra
]{revtex4-1}

\usepackage{graphicx}
\usepackage{dcolumn}
\usepackage{bm}
\usepackage{color}
\usepackage[breaklinks]{hyperref}

\def\id{{\mathbb I}}

\newcommand{\ket}[1]{|{#1}\rangle}
\newcommand{\bra}[1]{\langle{#1}|}

\begin{document}

\title{Experimental many-pairs nonlocality}%

\author{Hou Shun Poh}
\affiliation{Centre for Quantum Technologies, National University of Singapore, 3 Science Drive 2, Singapore 117543}

\author{Alessandro Cer\`{e}}
\affiliation{Centre for Quantum Technologies, National University of Singapore, 3 Science Drive 2, Singapore 117543}

\author{Jean-Daniel Bancal}
\affiliation{Department of Physics, University of Basel, Klingelbergstrasse 82, 4056 Basel, Switzerland}

\author{Yu Cai}
\affiliation{Centre for Quantum Technologies, National University of Singapore, 3 Science Drive 2, Singapore 117543}

\author{Nicolas Sangouard}
\affiliation{Department of Physics, University of Basel, Klingelbergstrasse 82, 4056 Basel, Switzerland}

\author{Valerio Scarani}
\affiliation{Centre for Quantum Technologies, National University of Singapore, 3 Science Drive 2, Singapore 117543}
\affiliation{Department of Physics, National University of Singapore, 2 Science Drive 3, Singapore 117551}

\author{Christian Kurtsiefer}
\affiliation{Centre for Quantum Technologies, National University of Singapore, 3 Science Drive 2, Singapore 117543}
\affiliation{Department of Physics, National University of Singapore, 2 Science Drive 3, Singapore 117551}

\date{\today}

\begin{abstract}

Collective measurements on large quantum systems together with a majority voting strategy can lead to a violation of the CHSH Bell inequality. In presence of many entangled pairs, this violation decreases quickly with the number of pairs, and vanishes for some critical pair number that is a function of the noise present in the system. Here, we show that a different binning strategy can lead to a more substantial Bell violation when the noise is sufficiently small. Given the relation between the critical pair number and the source noise, we then present an experiment where the critical pair number is used to quantify the quality of a high visibility photon pair source. Our results demonstrate nonlocal correlations using collective measurements operating on clusters of more than 40 photon pairs.

\end{abstract}

\maketitle


\section{Introduction}

The ability of detecting single quanta, already developed for some decades, is a crucial feature of experimental quantum technologies, and the whole thinking in quantum information science usually relies on it~\cite{aspect}. This notwithstanding, recent studies have considered situations in which single-quanta control and detection is \textit{not} available. For instance, in many-body systems measurements are performed collectively -- the same measurement is applied to all particles and the outcome produced is extensive in the system size -- so single quanta identification is lost. It is also common in such systems to have only access to few-body correlators, in which case single-quanta resolution is also lost~\cite{Pezze16}. Another example where single-quanta detection is not available is when quantum light is detected by biological systems~\cite{Brunner08,Sekatski09,Vivoli16,Dodel16}.

Prompted by interest in these systems, it is relevant to study what happens to the violation of Bell's inequalities. Several restrictions have been highlighted in the limit of large numbers of particles. For instance, Bell inequalities can't be violated if only few-body collective observables are measured~\cite{dago2011}, unless one adds assumptions~\cite{Tura14,Schmied16}. In a many-pair scenario, high-order collective measurements are also unable to lead to a Bell violation as soon as some realistic coarse-graining is present~\cite{NW09}. At the same time, it is also known that the ability to address single quanta is not necessary for violating a Bell inequality where $n$ particles are subjected to collective measurement processed through majority voting~\cite{bancal08}. In this case, however, the observed violation is known to decrease quickly as a function of the number of particles.

In this paper, we show that substantial violation can be obtained in presence of collective measurements for an arbitrary number of particles by using a parity binning strategy. We discuss the resistance to noise of this Bell violation as a function of the number of measured particles $n$ and compare it with the one obtained in the previous approach. In each case we find that the maximal cluster size $n_c$ for which a Bell violation can be obtained is sensitive to experimental imperfections and proves to be a good figure of merit to certify the quality of a high visibility source~\cite{Poh2015, Christensen:2015bl, Wang16}.
From this insight, we perform a proof-of-principle experiment using a very high quality source of photon pairs and demonstrate non-local correlations with collective measurements operating on clusters of up to 41 photon pairs.

\section{Theory}

\subsection{The many-pair scenario}

Consider a source that produces~$n$ independent pairs of correlated particles -- in particular, particles belonging to different pairs are \textit{a priori} distinguishable~\cite{bancal08}. One particle of each pair is sent to party Alice, the other to party Bob. Each party submits all its~$n$ particles to the same single-particle measurement, labeled~$x$ for Alice and~$y$ for Bob. Alice's (Bob's) particle from the~$i$-th pair returns the outcomes~$a_{xy}^i$ ($b_{xy}^i$).

We focus on the case where each party performs two measurements ($x,y\in\{1,2\}$) and the single-particle outcome is binary ($a,b\in\{0,1\}$).
The correlations observed in this scenario are nonlocal if and only if they violate a Bell inequality for two inputs and~$2^n$ outputs. For a given correlation, locality can be checked by a linear program, but the hope of completely solving the local polytope for large~$n$ is slim, since the full list of inequalities is already unknown for $n=2$~\cite{Bancal10, faacets}. The number of “liftings” (that is, loosely speaking, the number of different versions) of CHSH alone is exponential in~$2^n$.

We consider a family of measurements indexed by a single angle $\beta$ as follows: 
\begin{align}\label{eq:basis} A_1 &= \sigma_z, & A_2 &= \cos(2\beta)\sigma_z + \sin(2\beta)\sigma_x\nonumber\\ B_1 &= \cos\beta\ \sigma_z + \sin\beta\ \sigma_x, & B_2 &= \cos\beta\ \sigma_z - \sin\beta\ \sigma_x. \end{align}
When applied to the Werner state
\begin{equation}\label{eq:wernerstate}
    \rho=V\ket{\psi^-}\bra{\psi^-}+(1-V)\id/4\,,
\end{equation}
where $\ket{\psi^-}=\frac{1}{\sqrt{2}}(\ket{01}-\ket{10})$ is the maximally entangled state of two qubits,
the statistics of a single pair are described by the correlators 
\begin{equation}\label{eq:1box}
    E_{11}=E_{12}=E_{21}=V\cos\beta\;,\;E_{22}=V\cos(3\beta)
\end{equation}
where $E_{xy}=\textrm{Prob}(a_{xy}=b_{xy})-\textrm{Prob}(a_{xy}\neq b_{xy})$, and uniformly random marginals.

So far, no assumption has been made, but now we assume that each party is not able to observe the entire string of outcomes, but only their sum:
\begin{align}\label{manyboxdef}
    a_{xy}=\sum_{i=1}^n a_{xy}^i& \;,\;\; b_{xy}=\sum_{i=1}^n b_{xy}^i
\end{align}
with  $a_{xy},b_{xy}\in\{0,1,...,n\}$.
In other words, in a Stern-Gerlach picture, each party can count how many particles take each port, but is unable to sort out which of her particles was correlated with which of the other party's.

To simplify the test for Bell violation, we introduce a processing of the data so that
$a_{xy}\rightarrow a'_{xy}$ and $b_{xy}\rightarrow b'_{xy}$, with $a'_{xy},b'_{xy}\in\{+1,-1\}$,
bringing us back to a two-input and two-output scenario, in which the only relevant Bell inequality is the CHSH inequality
\begin{equation}\label{eq:CHSH_ineq}
    S_n = E_{11}^{(n)} + E_{12}^{(n)} + E_{21}^{(n)} - E_{22}^{(n)}\leq 2\,,
\end{equation} where $E_{xy}^{(n)}=\textrm{Prob}(a'_{xy}=b'_{xy})-\textrm{Prob}(a'_{xy}\neq b'_{xy})$.
If the correlations of the primed variables violate CHSH, certainly those of the original unprimed variables violated some Bell inequality (surely the corresponding lifting of CHSH~\cite{Pironio05}). Of course, information has been lost in the binning, so the converse is not true.

Specifically, we consider two such local binnings, majority vote and parity. For each of them, we estimate a lower bound on the Werner state visibility, $V$ as a function of the number of pairs $n$ at which a violation is observed.

\subsection{Majority vote}\label{sec:majority_binning}
The first binning, \textit{majority vote}, is obtained by comparing the observed output to a fixed threshold $t=n/2$. If the outcome is larger than $t$, we produce `+1', otherwise we produce `-1', i.e.
\begin{equation}\label{eq:majority_binning}
    a_{xy}' = \text{sign}(a_{xy}-t).
\end{equation}
Previous numerical studies suggest that the violation (with optimized measurement setting) of CHSH after such binning decreases roughly as $\sim 1/\sqrt{n}$ when $n$ is growing. 
For $n\lesssim 65$, one may numerically compute the minimal visibility for each $n$ for which violation is possible, 
\begin{equation}\label{Vcmaj}
    V^\text{maj}(n) \simeq 1-\frac{0.5690}{n} + \frac{0.2763}{n^2}.
\end{equation}
For instance, a violation with $n=21$ pairs of Werner states requires a visibility $V\geq 97.35\%$ \cite{kwiat:95}; a visibility $V\geq 99.12\%$ still achieves a violation until $n=64$ pairs.

\subsection{Parity binning}
Let us now consider the parity binning:
\begin{equation}\label{eq:parity_binning}
a'_{xy}=(-1)^{a_{xy}}
\end{equation} and similarly for Bob. Recalling \eqref{manyboxdef}, the bipartite correlator $E_{xy}^{(n)}=\langle a'_{xy}b'_{xy}\rangle$ is
\begin{equation}\label{eq:bipartite_corr}
\begin{split}
E_{xy}^{(n)} &= \langle (-1)^{\sum_i a_{xy}^i} \times (-1)^{\sum_i b_{xy}^i} \rangle \\
&= \langle \prod_i (-1)^{a_{xy}^i + b_{xy}^i} \rangle=(E_{xy})^n.
\end{split}
\end{equation} Remarkably, in absence of noise, the CHSH inequality can be significantly violated for arbitrarily large $n$. Indeed, the single-box correlators \eqref{eq:1box} for $V=1$ lead to
\begin{equation}\label{eq:chshpar}
    S_n=3\cos^n\beta - \cos^n(3\beta).
\end{equation}
Choosing $\beta=\frac{\beta_0}{\sqrt{n}}$, we find $S_n \xrightarrow{n\to\infty} 3e^{-\beta_0^2/2} -e^{-9\beta_0^2/2}$ whose maximum is $S_\infty=8\cdot 3^{-9/8} \simeq 2.32$ obtained for $\beta_0 = \sqrt{\ln 3}/2 \simeq 0.524$. 

This asymptotic violation $S_{\infty}>2$ disappears with the least amount of white noise, since $S_n(V)=V^nS_n(V=1)\xrightarrow{n\to\infty} 0$ for any $V<1$. Nevertheless, for every $n$ there exist a critical visibility $V_c(n)$, such that violation will be observed if $V>V_c(n)$. The condition $S_n\simeq 8\cdot3^{-9/8}V^n\simeq 8\cdot3^{-9/8} (1-n(1-V))=2$ gives 
\begin{equation}\label{Vcpar}V^\text{parity}(n) \simeq 1-\frac{1-3^{9/8}/4}{n} \simeq 1-\frac{0.14}{n}.\end{equation} 
This expression, as opposed to Eq.~\eqref{Vcmaj}, is not a numerical guess, but an analytic approximation in the high visibility regime. 
A violation with $n=4$ pairs requires a visibility higher than $V\geq 97\%$; a visibility $V\geq 99\%$ produces a violation with at least $n=14$ pairs.

Comparing parity binning with majority vote, we have noticed that the latter tolerates smaller values of $V$ insofar as the possibility of violation is concerned. However, the amount of violation is different in both cases: for majority vote, the violation quickly decreases with the number of pairs as $\sim1/\sqrt{n}$, whereas it only decreases linearly $\sim V_0-\beta n$ in the parity case. Therefore, for $V$ high enough, parity may exhibit higher violations for the same values of $n$. This behaviour starts at $V\gtrsim 99.4\%$ (see Appendix A, Fig.~\ref{fig:comparetheory}).

\section{Experiment}
\subsection{Experimental setup}
  \begin{figure}
    \includegraphics[width=.75\columnwidth]{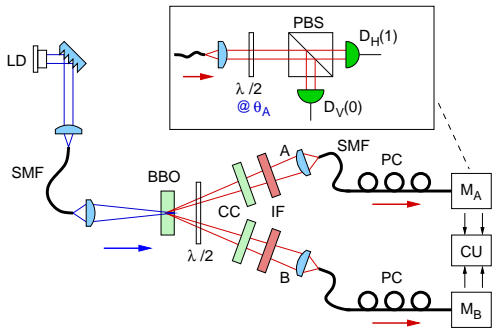}
    \caption{Schematic of the experimental set-up.
    Polarization correlations of entangled-photon
    pairs are measured by the polarization analyzers   M$_A$ and M$_B$,   each
    consisting of a half wave plate ($\lambda/2$) followed by a polarization beam
    splitter (PBS).
    All photons are detected by Avalanche photodetectors D$_H$ and
    D$_V$, and registered in a coincidence unit (CU).}
    \label{fig:setup}
  \end{figure}

In our experiment (see figure~\ref{fig:setup}), the output of a
grating-stabilized laser diode (LD, central wavelength 405\,nm) passes through
a single mode optical fiber (SMF) for spatial mode filtering, and is focused
to a beam waist of 80\,${\mu}$m into a 2\,mm thick BBO crystal cut for type-II
phase-matching. There, photon pairs are generated via spontaneous parametric
down-conversion (SPDC) in a  non-collinear configuration, with a half-wave
plate ($\lambda$/2) and a pair of compensation crystals (CC) to take care of
the temporal and transversal walk-off~\cite{kwiat:95}. Two spatial modes ($A$,
$B$) of down-converted light, defined by the SMFs for 810\,nm, are matched to
the pump mode to optimize the collection~\cite{kurtsiefer:01}. In type-II
SPDC, each down-converted pair consists of an ordinary and extraordinarily
polarized photon, corresponding to horizontal ($H$) and vertical ($V$)
polarization in our setup.
Polarization controllers (PC) minimize the polarization
transformation caused by the SMFs to the collected modes.

One of the CC is tilted to adjust the phase between the two decay possibilities,
obtaining an output state very close to the
singlet polarization Bell state
\mbox{$\ket{\psi}=1/\sqrt{2}\left(\ket{H}_A\ket{V}_B-\ket{V}_A\ket{H}_B\right)$}.

In the polarization analyzers (inset of figure~\ref{fig:setup}), photons from
SPDC are projected onto the linear polarizations necessary for the Bell tests
by $\lambda$/2 plates, set to half of the analyzing angles $\theta_{A(B)}$,
and polarization beam splitters with extinction ratios of 1/2000 and 1/200 for
transmitted and reflected arms. Photons are detected by avalanche photo diodes (APD), and corresponding detection events from the same pair identified by a coincidence unit if they arrive within $\approx\pm$3\,ns of each other.

The quality of polarization entanglement is assessed in the traditional way
via the polarization correlations in a basis complementary to the intrinsic HV
basis of the crystal. With interference filters (IF) of 5\,nm bandwidth (FWHM)
centered at 810\,nm, we observe a visibility $V_{45}$ = 98.68$\pm$0.20\%  in
the $45^\circ$ linear polarization basis. In the natural H/V basis of the
type-II down-conversion process, the visibility reaches $V_{\rm HV}$ =
99.67$\pm$0.12\%.

Non perfect symmetry of the collection modes can lead to ``colored'' noise, i.e. photon pairs that show anti-correlation only in a specific measurement basis~\cite{White:1999kd}, reducing the quality of the state.
In a previous experiment~\cite{Poh2015}, we have already estimated the very high quality of the state generated by this source. 
The non-ideal visibility is due to the non-perfect neutralization of the
polarization rotation caused by the SM fibers. This affects the outcome of
the violation observed, as we discuss more in details later.

\subsection{Measurement and Post-processing}

In this proof of principle experiment, we did not aim for a loophole-free
demonstration. Due to the limited efficiency of the APD detectors and the
source geometry, we assume that the detected photons are a fair sample of the
entire ensemble. Similarly, even though Alice and Bob are not space-like
separated, we assume that no communication happens between measurements on
both sides. Moreover, the basis choice is not random, as necessary for a Bell
test. Instead, we set the basis and record the number of events in a fixed
time. Based on or experience with the setup, we assume that the state generated by the source and all the
other parameters of the experiment do not change significantly between
experimental runs.

A single measurement run lasts $60\,$s, during which we record an average of
$16\times10^3$ coincidences between detectors at Alice and Bob. A detection
event at the transmitted output of each PBS is associated with 0, at the
reflected one with 1.
We discard any two-fold
coincidences between detectors belonging to the same party, corresponding to
multiple pairs of photos generated within the coincidence time window.
From the detected single rates, we calculate an expected
rate for these events of~$\approx8.9\times10^{-6}$~1/s.

To avoid a bias due to the asymmetries in detector efficiencies, we
record coincidences not only in a basis $(A_j,B_k)$, but also in three
equivalent bases $(A_j+45^\circ,B_k)$, $(A_j, B_k+45^\circ)$, and
$(A_j+45^\circ,B_k+45^\circ)$. A rotation by $45^\circ$ effectively swaps the
roles of the transmitted and reflected detectors. Each party, when using such
a rotated basis, needs to invert the measurement outcome. We repeat these
measurement sets for a range of $\beta$, and the corresponding four bases defined by Eq.~(\ref{eq:basis}).

To replicate the many-box scenario, we organize the sequence of results into
clusters of size $n$ for every set of measurement angles. For each cluster we
calculate the majority (parity) binning using Eq.~\eqref{eq:majority_binning}
(Eq.~\eqref{eq:bipartite_corr}). Following the procedure in~(\ref{eq:CHSH_ineq}), we obtain a value of
$S_n$ for every $n$ of interest. To evaluate the error associated to every
$S_n$, the same procedure is repeated~1000 times, shuffling the order of the
results every time before the clustering.

\subsection{Discussion}
\begin{figure}
  \begin{center}
    \includegraphics[width=\columnwidth]{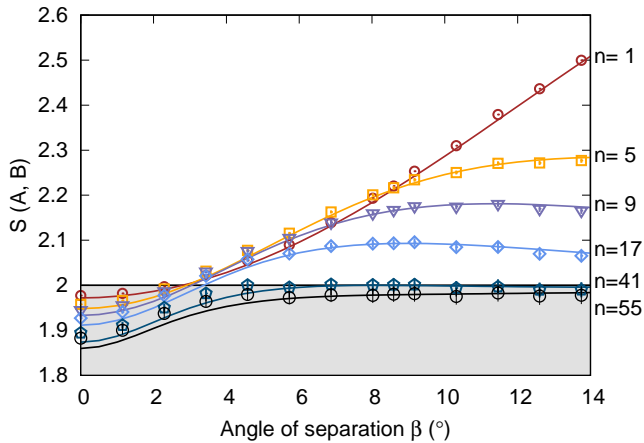}
   \caption{ \label{fig:exp_majority} 
   Majority processing for different $n$ applied to the data.
   The error bars are obtained from the bootstrapping procedure indicated in the text.
  The continuous lines are obtained numerically following section~\ref{sec:majority_binning}, with  $V\;=\;0.9892$.
   }
  \end{center}
\end{figure}

\begin{figure}
  \begin{center}
    \includegraphics[width=\columnwidth]{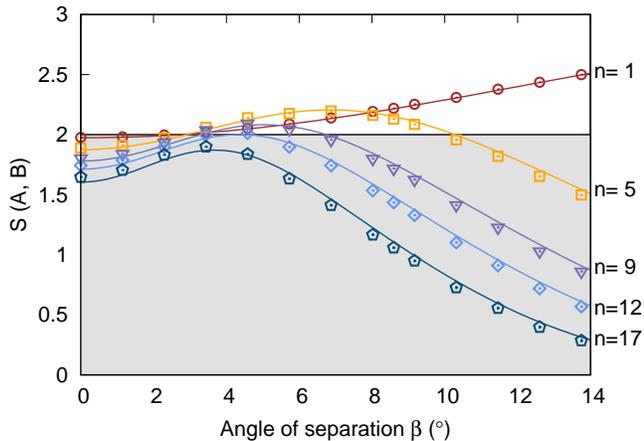}
   \caption{ \label{fig:exp_parity}
   Parity processing for different $n$ applied to the data. The error bars are obtained from the bootstrapping procedure indicated in the text. The continuous lines are calculated using Eq.~(\ref{eq:chshpar}) with $V$ = 0.9871.
    }
  \end{center}
\end{figure}

The results of the measurement are reported in fig.~\ref{fig:exp_majority} for the majority vote and fig.~\ref{fig:exp_parity} for the parity binning.
We estimate $n_c$ in both cases by identifying the largest $n$ that still shows a violation of inequality~(\ref{eq:CHSH_ineq})
For the case of majority vote, $n_c^\text{maj}=41$.
The continuous lines in fig.~\ref{fig:exp_majority} are obtained numerically, using as input a Werner state with $V=V_c^\text{maj}=0.9892$ (c.f. Eq.~\eqref{Vcmaj}).
Since the white noise of a Werner state corresponds to a worst-case scenario (any source with colored noise, with $V$ the minimal visibility over all choices of bases, will perform at least as well as the corresponding Werner state), the continuous lines are a lower bound on the observed violation. 
In fig.~\ref{fig:exp_parity} we observe that this is true indeed from small values of the angle $\beta$.
Instead, for larger angles the experimental violation is smaller than the predicted lower bound. This is due to a rotation of the measurement basis due to the imperfect neutralization of the SM fibers. Due to the specific alignment procedure, this rotation affects the detected visibility more for larger angles, as indicated by the relatively low  $V_{45}$ = 98.68$\pm$0.20\%  in
the $45^\circ$ linear polarization basis.
Reproducing the exact violation expected would require an extensive characterization of the rotation induced by the fibers that would not add to much to the present demonstration.

A similar procedure is applied to the parity binning.
In this case, we find $n_c^\text{parity}=12$.
The continuous lines of fig.~\ref{fig:exp_parity} were obtained using Eq.~(\ref{eq:chshpar}) with $V_c^\text{parity}=0.9871$ (c.f. Eq.~\eqref{Vcpar}).
Similar conclusions regarding the effect of the imperfect neutralization of the SM fibers can be drawn.

\section{Conclusion}
We considered a many-pair scenario, where $n$ identical entangled pairs are produced and measured collectively, and showed experimentally that a Bell inequality can be violated in this scenario. The maximal number of pairs for which a violation can be observed quantifies the high quality of the pair source. In our experiment we report a violation up to 41 pairs in presence of majority voting, and 12 pairs in presence of parity bining. We also prove analytically that a violation can be observed in presence of collective measurement for any number pairs $n$, and that this violation can remain significant for arbitrary $n$ in the noiseless limit.

\section*{Acknowledgments.} We thank Enky Oudot
for feedback and discussions. This research is supported by the Singapore
Ministry of Education Academic Research Fund Tier 3 (Grant
No. MOE2012-T3-1-009); by the National Research Foundation, Prime Minister's office, and the Ministry of Education, Singapore, under the Research Centres of Excellence programme; by the Swiss National Science Foundation (SNSF), through the NCCR QSIT and the Grant number PP00P2-150579; and by the John Templeton Foundation Grants 59342 ``Is the human eye able to see entanglement?'' and 60607 ``Many-box locality as a physical principle''.

%

\clearpage

\begin{appendix}

\onecolumngrid
\section{Amount of Bell violation with parity binning}

In the main text we discuss the relation between the number of pairs at which a Bell violation can still be observed, for either majority of parity binning, and the quality of the source in terms of visibility $V$. The amount of Bell violation that is obtained in the many-pair scenario when using a majority binning is described in~\cite{bancal08}. Here we analyse how the amount of Bell violation depends on the number of pairs in the case of parity binning and compare it to the majority case. In particular, we show that its decreases more and more slowly as the visibility increases.

To see this, we consider the CHSH expression~\eqref{eq:chshpar}, together with the choice of setting
\begin{equation}
\beta=\frac{\beta_0}{\sqrt{n}}\ ,\ \ \beta_0=\frac{\sqrt{\ln(3)}}{2}.
\end{equation}
As discussed in the main text, these settings give rise to a violation for a number of pairs smaller than
\begin{equation}
n_c(V)=\frac{1-3^{9/8}/4}{1-V}.
\end{equation}

\begin{figure}
  \begin{center}
    \includegraphics[width=.4\columnwidth]{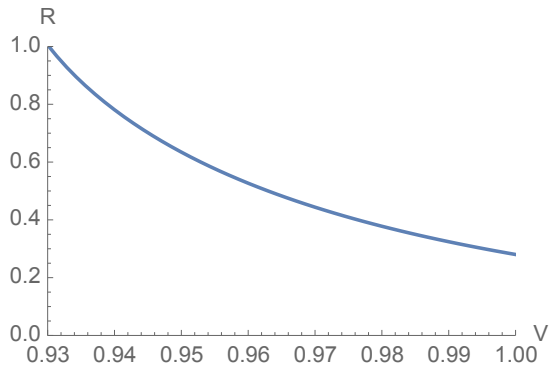}

   \caption{ \label{fig:parityRatio}
   Amount of Bell violation remaining in the parity case when considering $n=n_c/2$ pairs.
    }
  \end{center}
\end{figure}

We then estimate the sensibility of the Bell violation to the number of pairs by computing the amount of violation that can still be observed when the number of pairs is half of the maximum possible number, i.e. $n=n_c/2$. For this, we define the ratio
\begin{equation}
R = \frac{S^n(V,n=n_c(V)/2) - 2}{S^n(V,n=1) - 2}.
\end{equation}
This quantity is represented in figure~\ref{fig:parityRatio}. Interestingly, only a fraction of the initial violation is lost independently of the visibility. The decrease in violation is thus linear in $n$.

Moreover, since the number of pairs considered here increases with the visibility, the Bell violation with parity binning becomes less and less sensitive to the number of pairs as the visibility increases. This contrasts with the case of majority voting, where the violation is upper-bounded by the case $V=1$, which decays as $\sim1/\sqrt{n}$.

\begin{figure}
  \begin{center}
    \includegraphics[width=0.8\columnwidth]{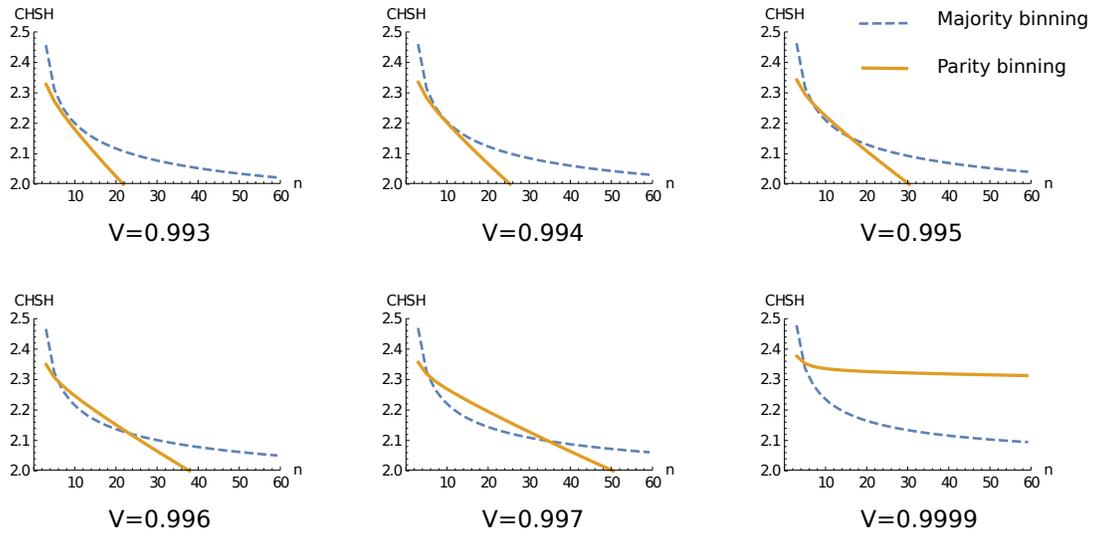}
   \caption{ \label{fig:comparetheory}
   CHSH violation achieved by the majority and parity binnings as a function of the source visibility $V$ and number of pairs $n$. For $V<0.994$, the largest Bell violation is achieved by the majority strategy. For $V>0.994$, the parity strategy provides a large violation for a range of $n$.
    }
  \end{center}
\end{figure}

Given this qualitative difference between the Bell violation provided by the majority and parity binnings, one should expect that the Bell violation provided by the parity binning would outperform the one provided by the majority procedure for a sufficiently large visibility. From figure~\ref{fig:comparetheory}, we see that this cross-over occurs around $V=0.994$.

\end{appendix}

\end{document}